\documentclass[preprint,showkeys]{revtex4}
\usepackage{graphicx}
\usepackage[english]{babel}
\usepackage[utf8]{inputenc}
\usepackage{subfig}
\usepackage{floatrow}
\usepackage{comment}
\usepackage{amsmath}
\usepackage{amssymb}

\begin{document}

\title{A Wide Optical-Gap in Fully sp$^3$-Like Hydrogenated Monolayer Graphene}

\author{Alice Apponi$^{1,2}$}
\author{Orlando Castellano$^{1,2}$}
\author{Daniele Paoloni$^{1,2}$}
\author{Domenica Convertino$^3$} 
\author{Neeraj Mishra$^{3}$}
\author{Camilla Coletti$^{3,4}$} 
\author{Carlo Mariani$^{5,6}$}
\author{Alessandro Ruocco\footnote{Corresponding author. E-mail: alessandro.ruocco@uniroma3.it}$^{1,2}$}

\affiliation{$^1$Dipartimento di Scienze, Universit\`a degli Studi di Roma Tre, Via della Vasca Navale 84, 00146 Roma, Italy}
\affiliation{$^2$INFN Sezione di Roma Tre, Via della Vasca Navale 84, 00146 Roma, Italy}
\affiliation{$^3$Center for Nanotechnology Innovation @NEST, Istituto Italiano di Tecnologia, Pisa, Italy}
\affiliation{$^4$Graphene Labs, Istituto italiano di tecnologia, Via Morego 30, I-16163 Genova, Italy}
\affiliation{$^5$Sapienza Universit\`a di Roma, Piazzale Aldo Moro 2, 00185 Roma, Italy}
\affiliation{$^6$INFN Sezione di Roma, Piazzale Aldo Moro 2, 00185 Roma, Italy}

\begin{abstract}
A comprehensive spectroscopic characterisation of two highly hydrogenated monolayer graphene samples transferred onto nickel grids is reported. With X ray photoemission spectroscopy on the C 1s core-level, a 100$\%$ $sp^3$ profile was observed upon hydrogenation of a more $sp^3$-like initially defected graphene, while a flatter, more $sp^2$-arranged, graphene reached a 62$\%$ $sp^3$ saturation. Low-energy reflection electron energy-loss spectroscopy (EELS) corroborates these findings through the $\pi$-plasmon excitation quenching for the fully $sp^3$ sample and a significant reduction for the partially converted one. The extreme surface sensitivity of low-energy reflection EELS enables extraction of the optical band gap of the hydrogenated layer even on a metallic support, yielding values of 6.3 and 6.2 eV for the two samples. The C--H stretching vibrational mode is also resolved, providing a direct fingerprint of graphene--hydrogen bonding. Finally, valence-band measurements of the $62\%$ saturated sample suggest the coexistence of one-sided and two-sided hydrogenation morphologies.
\end{abstract}


\maketitle

\section{Introduction}
\label{sec:intro}
Graphene is an atomically thin carbon sheet, which aroused great interest ever since it was isolated from graphite \cite{Novoselov1}. The 2D nature of this material accompanied by its unique mechanical, thermal and electronic properties \cite{Novoselov, PAPAGEORGIOU} make graphene attractive for many applications. Moreover, the chemical modification of graphene, and in particular its functionalisation with hydrogen, has enhanced the attractiveness of this material by opening the way to even further applications. Among them, the storage of hydrogen for energy and fuel-cells \cite{Dillon} or for applications in particle physics experiments, as the PTOLEMY project \cite{Cocco, PTOLEMY4, Apponi_2022}. In the latter, graphene structures are thought to be employed in the target for the storage of tritium, the $\beta$-emitter from the decay of which the neutrino mass should be measured. Since tritium is a radioactive isotope that requires specific facilities for handling, the hydrogenation of graphene represents the reference study. 

Furthermore, in the field of electronic devices, the application of graphene is limited by its characteristic to be a zero-gap semiconductor, with the band touching at the Dirac points. The adsorption of hydrogen on graphene leads instead to the modification of the pristine electronic structure by breaking the $\pi$-bonds, thus producing a transition from $sp^2$- to $sp^3$- hybridised carbon atoms and the opening of a band gap \cite{Boukhvalov, Sofo}. The versatility of this material, and its growing list of potential applications, possibly made hydrogenated graphene the most studied functionalisation of carbon nanostructures. Many efforts have been made both experimentally and theoretically for the synthesis, the characterisation and the comprehension of the properties of partially and fully hydrogenated graphene (usually named \emph{graphane}). Nevertheless, the scenario is fragmented and still presents many open questions.

From the experimental point of view, the hydrogenation has been investigated for a plethora of graphene synthesis procedures and substrates: graphene grown on Cu via chemical vapor deposition (CVD) \cite{Luo}, grown CVD and then transferred onto SiO$_2$/Si substrates \cite{Burgess, Zhao}, mechanically exfoliated from graphite on SiO$_2$/Si \cite{Elias, Ryu, Luo2, Felten}, graphene grown by ethylene pyrolysis on Ir(111) \cite{Balog} or epitaxially on Pt(111) \cite{Panahi}. Furthermore, there are experiments based on different forms of graphene, such as quasi free-standing realisations obtained by the Au intercalation into graphene CVD grown on Ni \cite{Paris, Haberer} or even free-standing nanoporous graphene \cite{NPG21, NPG22}. The most common hydrogenation techniques are based on the exposure to either cold or hot hydrogen plasma, with atomic H produced by the thermal cracking of H$_2$ gas or via chemical reactions like the Birch reduction.

The mechanism of hydrogen bonding to graphene, a planar honeycomb lattice, is to pull one of the carbon atoms out of plane, leading to the deformation of the $sp^2$ configuration of carbon atoms toward an $sp^3$-like coordination\cite{Boukhvalov}. Hydrogen adsorption seems also to be favoured by the higher $sp^3$-like character or convex $sp^2$ configurations of curved graphene nanostructures, which lowers the barrier for a C atom to be pulled out above the plane defined by its nearest neighbours \cite{Ruffieux, Park, Heisenberg}. In addition, it is worth noting that $sp^3$ defects in the pristine graphene lattice represent highly reactive and unstable sites. Indeed, the dangling bond of $sp^3$ coordinated carbon easily reacts with atoms like hydrogen, oxygen or other functional groups \cite{ENOKI, OLIVEIRA, IACOBUCCI}. Therefore, both in ultra-high vacuum, where the residual gas has poor oxygen and significant hydrogen contents, and during hydrogenation processes, the $sp^3$ defects in graphene are either dangling or -- most likely -- passivated with hydrogen terminations. 

The highest H uptake on graphene reported so far has been achieved in 2022 by Betti \emph{et al.} for a sample of nanoporous graphene hydrogenated with thermally-cracked H atoms \cite{NPG22}. The loading has been investigated with nano-X-ray photoemission spectroscopy on the C 1s core-level. The $sp^3$ intensity over the total one was 90$\%$ at hydrogen saturation.

The morphology and the realisation of the C-H bonding in hydrogenated graphene is another open discussion, on both the theoretical and experimental level. The question is made even more intriguing by the fact that the band gap is expected to depend on the morphology and coverage, thus it might be tuned \cite{Sofo, Gao, Lebegue, Leenaerts, Zhou}. Indeed, all the possible configurations of half or fully hydrogenated graphene differ in calculated band structures and (optical) band gaps. Fully hydrogenated graphene is also expected to be a wide gap semiconductor, although the gap energy is yet to be univocally determined.  The predicted gap energies for partially hydrogenated graphene and graphane range between 4.7 eV and 6.1 eV \cite{NPG22, Gao, Lebegue, Cudazzo, Betti23}. The measurement of the optical band gap via photon absorption or electron energy loss spectroscopy, aside from all the technical and experimental difficulties, may provide a clue to solve the theoretical puzzle. On the other hand, also the investigation of the (valence) band structure, through UV-photoemission spectroscopy, is a promising experimental test for the theoretical predictions. 

Before entering into the details of the experimental results, it is useful to summarise the expected effects of the hydrogen bonding to a graphene-like carbon structure: change of the carbon atoms configuration from $sp^2$ to $sp^3$-like; quenching of the $\pi$-plasmon collective excitation, associated to extended domains of $sp^2$ coordinated carbon atoms, and the opening of a band gap. 
The above mentioned effects can be spectroscopically observed as \emph{indirect} footprints of the H adsorption on the graphene lattice. In addition to these, a \emph{direct} evidence of the C-H bonding formation can be also measured, with proper energy resolution, via electron energy loss spectroscopy: the C-H stretching mode.
 
All these hydrogenation effects were studied \emph{in-situ} on the two samples of monolayer graphene as a function of the H dose, by combining three spectroscopic techniques: X-ray photoemission spectroscopy (XPS), electron energy loss spectroscopy (EELS) and UV photoemission spectroscopy (UPS). In light of the studies recently reported on the stability of hydrogenated graphene in vacuum and in air \cite{stability}, the fact that the hydrogenation process and the spectroscopic characterisations were performed \emph{in-situ}, without breaking the ultra-high-vacuum conditions, deserves particular attention. Indeed, exposure of hydrogenated graphene to air leads to rapid oxidation and, consequently, degradation of the sample, whereas long-term stability was observed when the sample was kept under ultra-high-vacuum conditions.

\section{Sample preparation and experimental methods}
\label{sec:prep}
The sample preparation is carried out at CNI@NEST laboratory in Pisa where polycrystalline monolayer graphene is grown via chemical vapor deposition (CVD) on electropolished copper \cite{Miseikis_2015, Convertino2020} and then transferred onto a transmission electron microscopy (TEM) nickel grid with the standard wet etching technique \cite{Miseikis_2015, Li2009, D1CP04316A}. The grids employed are the commercially available G2000HAN by Ted Pella Inc. (nominal hole diameter 6.5 $\mu$m and pitch 12.5 $\mu$m) made of nickel only, without any additional film or mesh. Further details on this procedure are described in \cite{Apponi2024}. After preparation, the samples are inserted in the ultra-high vacuum (UHV) chamber of the LASEC laboratory in Roma Tre University, where a 550 $^\circ$C annealing is performed in order to clean the graphene from PMMA (polymethyl methacrylate) residues due to the transfer procedure \cite{Apponi2024}. Two monolayer graphene on TEM samples were studied for the hydrogenation, both firstly annealed at 550 $^\circ$C, called samples A and B in the following. These two graphene on TEM samples are the very same samples A and B employed for the electron transmission measurements reported in \cite{Apponi2024}. As thoroughly investigated in the paper, the cleaning treatment leads to the partial breaking of the graphene suspended on the holes of the TEM grid. Indeed, the portion of suspended graphene - over the total - after annealing is about 14$\%$ and 20$\%$ for samples A and B respectively \cite{Apponi2024}. The graphene hydrogenation is carried out \emph{in-situ} by means of a FOCUS EFM-H atomic hydrogen source, with hydrogen partial pressure of 3.6 $10^{-6}$ mbar and heating power 25 W (capillary temperature about 2100 K). 

The UHV chamber (base pressure $1 \cdot 10^{-10}$ mbar) is also equipped with the experimental apparatuses for XPS, UPS and EELS. For all the spectroscopies, a hemispherical electron analyser (66 mm radius) with a position sensitive detector for parallel acquisition is employed. The XPS is performed with an Omicron XM1000 monochromatised Al K$\alpha$ X-ray source (h$\nu$ = 1486.7 eV) and the total energy resolution for the XPS measurements is 460 meV. The binding energy scale has been calibrated with a clean sample of highly oriented pyrolytic graphite (HOPG) by setting the binding energy of the C1s core-level at 284.5 eV \cite{Chen}. A He discharge lamp (FOCUS HIS 14 HD 250) with nominal spot size of 370 $\mu$m x 364 $\mu$m has been employed for UPS. The total energy resolution for this technique is 350 meV, with HeI$_\alpha$ (21.22 eV) photon energy. Finally, the source of electrons for EELS is a custom-made monochromatic electron gun (45 meV source energy resolution) operated at fixed electron energy of 91 eV \cite{Apponi_2021} and the total energy resolution for EELS is 60 meV. The electron scattering is performed in reflection geometry and the beam size on sample is approximately 500 $\mu$m.

\section{Results and discussion}

\subsection*{X-ray photoemission spectroscopy}
\label{sec:GrTEMH_XPS}
\begin{figure}[h]
\centering
\includegraphics[width=0.8\textwidth,height=0.6\textheight,keepaspectratio,trim={3cm 3cm 4cm 3cm},clip]{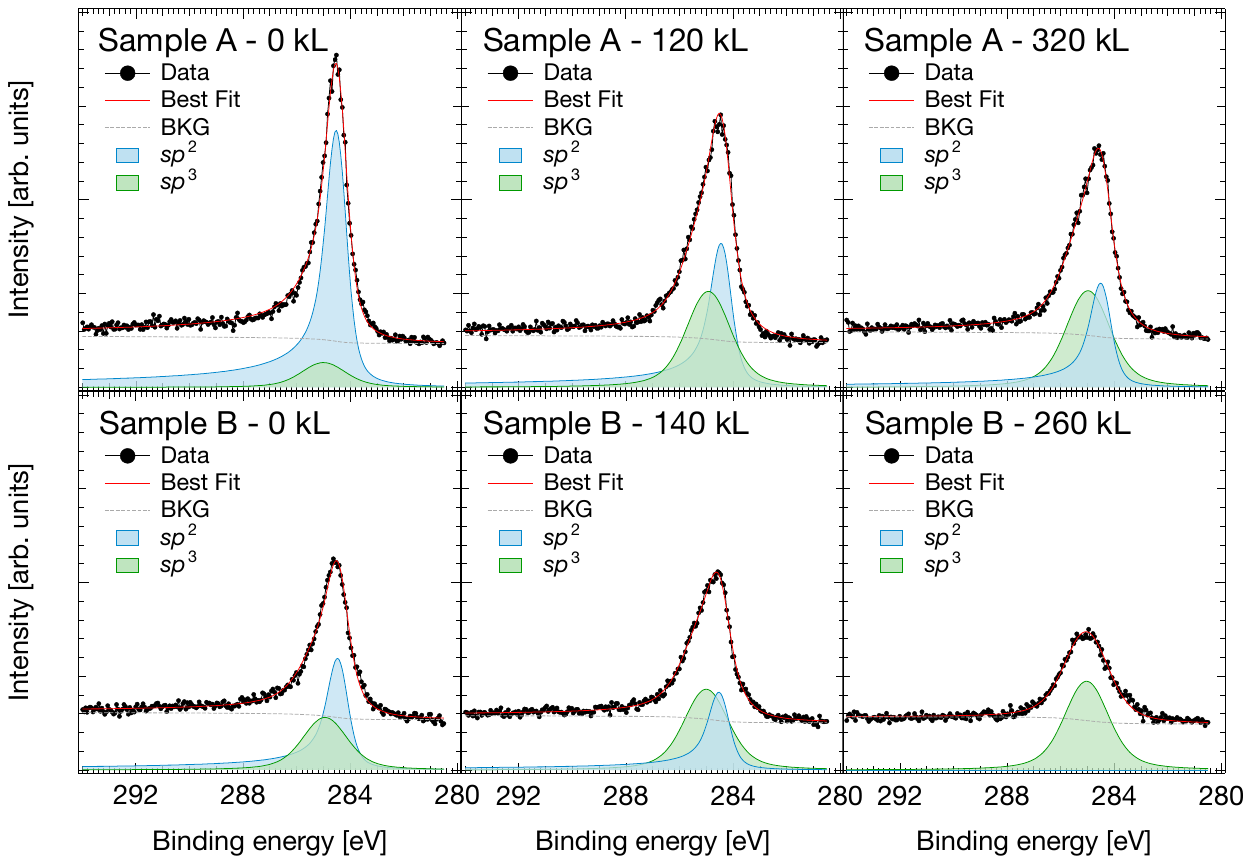} 
\caption{XPS C 1s spectra are reported on the top panel for sample A acquired at 0 kL,120 kL and 320 kL (left-to-right), on the bottom panel for sample B at 0 kL,140 kL and 260 kL (left-to-right). The black dots represent the experimental data, the red solid line is the sum of the fitting curves, the grey dashed line is the background and the fit profiles follow the colour code reported in the legends.}
\label{fig:GrTEMH_XPS}
\end{figure}

Measurements of the C 1s core-level were performed with XPS on samples A and B after the 550 $^\circ$C annealing treatment and after increasing exposure to atomic H. In Figure \ref{fig:GrTEMH_XPS}, the evolution of the C 1s spectrum as a function of the H dose is shown for the two samples. The analysis, reported for each spectrum, was carried out with a global fitting procedure. This method consists in the fit of the spectra all at once, by using two type of parameters: the \emph{global} parameters which are unique for all the spectra and the \emph{individual} parameters which are instead attributed to the single spectrum (more details on this procedure can be found in \cite{PAOLONI23}). In this case, the analysis was carried out on the spectra relative to the two samples at each hydrogenation step, with a single global fit. For the $sp^2$ component, all the Doniach-Sunjic line-shape parameters (asymmetry, Gaussian width and Lorentzian width) were free global parameters. On the other hand, the $sp^3$ component was fitted with a symmetric profile. The binding energy shift, with respect to the $sp^2$ position, the Gaussian and Lorentzian widths were free global parameters. The free individual parameters were instead the area of the components and the $sp^2$ binding energy. The latter was found at 284.45 eV in all the spectra, with a variability of 50 meV at most. The results for the global parameters are reported in Table \ref{tab:globalPar}.

\begin{table}[h]
\centering
\begin{tabular}{cccccc}
\multicolumn{5}{c}{Global parameters} \\
\hline \hline
Component \ & Energy shift [eV] \ & GW [eV] \ & LW [eV] \ & Asymmetry \\ \hline
$sp^2$ & 0 & 0.59 & 0.47 & 0.08  \\ \hline
$sp^3$ & 0.54  & 1.06 & 1.23  & 0 \\ \hline \hline
\end{tabular}
\caption{Global parameters resulted from the fit analysis on the C 1s core-level measured for samples A and B at increasing hydrogen dose. Component, binding energy shift with respect to the $sp^2$ position ($\pm$ 0.01), Gaussian width (GW) ($\pm$ 0.03 ), Lorentzian width (LW) ($\pm$ 0.03 ) and peak asymmetry ($\pm$ 0.01) are listed.}
\label{tab:globalPar}
\end{table}

The two samples resulted to be very different after preparation: sample A has a $sp^3$/($sp^2$+$sp^3$) area ratio of about 20$\%$, while for sample B the ratio is significantly higher (46$\%$) indeed. This result indicates a more defective character of sample B. The $sp^3$/($sp^2$+$sp^3$) ratio then increases for both as expected, with increasing hydrogen dose. However, sample B reaches a (100 $\pm$ 2)$\%$ $sp^3$ saturation after 260 kL, while for sample A the $sp^2$ component is still present also after 320 kL. In addition, an overall loss of 20 and 30 $\%$ in the C 1s total intensity was also observed for sample A and B respectively.

\begin{figure}[h]
\centering
\includegraphics[width=0.7\textwidth,height=0.6\textheight,keepaspectratio,trim={1cm 4cm 2cm 4cm},clip]{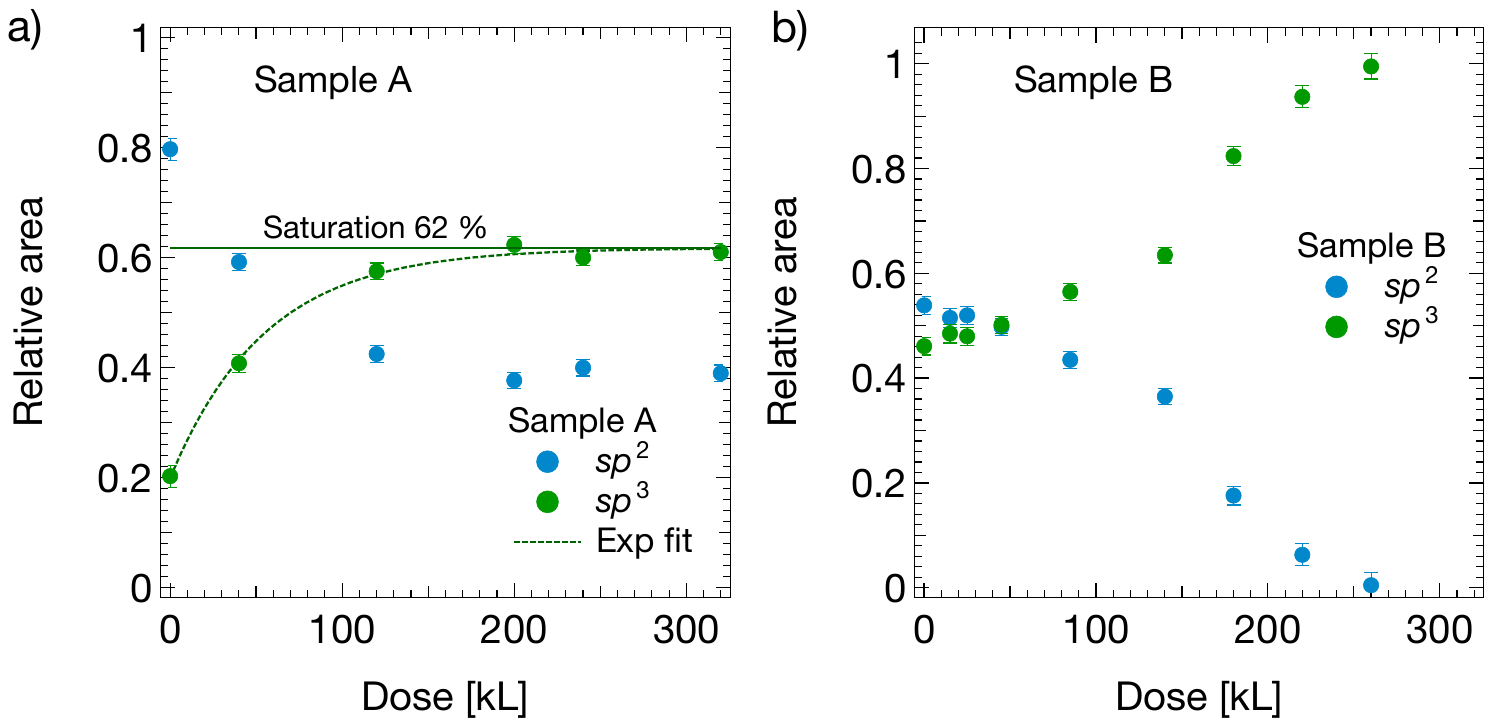} 
\caption{Area ratio $sp^2$/($sp^2$+$sp^3$) (blue) and $sp^3$/($sp^2$+$sp^3$) (green) as a function of the H dose for sample A (a) and for sample B (b). The green dashed line in (a) represents the exponential fit of the curve associated to the $sp^3$ area and the green solid line the saturation level resulted from the fit.}
\label{fig:GrTEMH_XPSarea}
\end{figure}

In Figure \ref{fig:GrTEMH_XPSarea}, the relative areas $sp^2$/($sp^2$+$sp^3$) and $sp^3$/($sp^2$+$sp^3$) are plotted as a function of the H dose for the two samples. The curve associated with the $sp^3$ area for sample A was also fitted with an exponential function and a saturation of (62 $\pm$ 1)$\%$ was found. What emerged from this XPS characterisation is that the hydrogen sticking seems to be favoured for graphene with higher $sp^3$-like concentration, like in the case of sample B. This suggests that an already distorted lattice is easier to be further distorted, while larger $sp^2$ domains seem to be more stable. This result is also in good agreement with the conclusions reported in \cite{Ruffieux}, as previously discussed.

According to the calculations and the experimental results reported in \cite{Betti23, Lizzit}, the specific morphology of the hydrogen bonding to the graphene lattice gives rise to different core-level shifts of the $sp^3$-like component, with respect to the $sp^2$. If a two-sided hydrogenation (C:H ratio 1:1 with H atoms bonded to C atoms on the two lattice sides alternatively) is achieved, a single $sp^3$ component should rise with a shift of 0.8 eV \cite{Betti23}. Conversely, for the one-sided hydrogenation (\emph{i.e.} H atoms bonded on one side only of the graphene lattice, thus a maximum H uptake of 50$\%$) two components shifted at 1.3 eV and -0.1 eV should be found \cite{Betti23}. On the other hand, only a one-sided hydrogenation was considered in \cite{Lizzit} and two shifted component, for high uptakes, were found at 0.17 eV and -1.04 eV. It should be underlined that, although the hydrogenation technique was the same in the two cases, the H uptake investigated in \cite{Betti23} was significantly higher.

In our experiment, the $sp^3$ binding energy shift was a global fit parameter and it resulted to be 0.54 eV, thus compatible with a two-sided hydrogenation. It should be also noticed that the width of this component is large (full width 1.6 eV), thus it could compensate the contribution of one-sided unresolved components. Therefore, XPS characterisation gives a clear indication of a two-sided hydrogenation, but the coexistence of one-sided hydrogenation cannot be excluded. Finally, it is important to underline that the $sp^3$/($sp^2$+$sp^3$) area ratio of 100$\%$ observed on sample B is the highest ever achieved among the experimental results on hydrogenated graphene-structures reported so far. 

\subsection*{Electron energy loss spectroscopy}
\label{sec:GrTEMH_EELS}
\begin{figure}[h]
\centering
\includegraphics[width=0.8\textwidth,height=0.6\textheight,keepaspectratio,trim={3.5cm 6cm 4.5cm 6cm},clip]{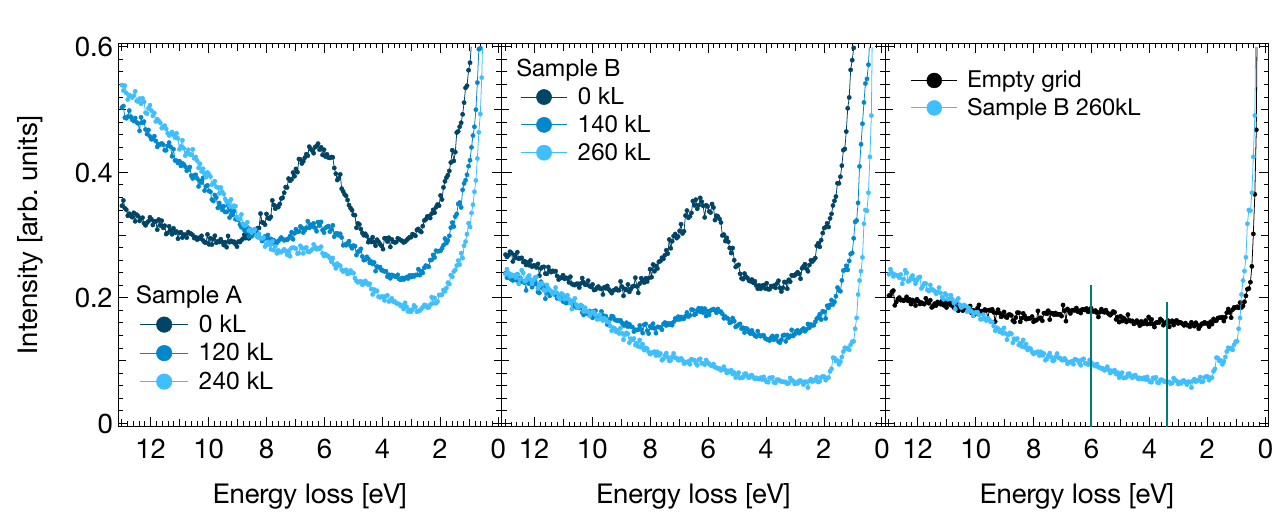} 
\caption{EELS spectra at increasing H dose are reported on the left for sample A acquired at 0 kL,120 kL and 240 kL (dark to light blue) and in the middle panel for sample B at 0 kL,140 kL and 260 kL (dark to light blue). On the right, a comparison between the EELS spectrum measured on an Ni TEM grid without graphene ("Empty grid") in black and on sample B after 260 kL of H dose in light blue is shown. Two lines at 6 and 3.4 eV were added as guides for the eye.}
\label{fig:GrTEMH_EELS}
\end{figure}

The behaviour of the $\pi$-plasmon excitation as a function of the H dose and the CH stretching mode were studied with the EELS technique. In Figure \ref{fig:GrTEMH_EELS}, the experimental data in the $\pi$-plasmon region are shown for sample A (0 kL,120 kL and 240 kL) and for sample B (0 kL,140 kL and 260 kL). The decrease of the peak around 6.2 eV, identified as the $\pi$-plasmon of graphene, is well visible as a function of hydrogen exposure for both samples (for a detailed study on the $\pi$-plasmon of clean graphene on grid see \cite{Apponi2024}). However, analogously to what resulted from XPS measurements, the evolution of the spectra shows significant differences in the two cases. For sample A, a substantial decrease in the intensity was observed as a function of the H dose but not its complete quenching. This result is compatible with the XPS outcome, which revealed a non-vanishing contribution of $sp^2$. 
Conversely, for sample B, the $\pi$-plasmon peak results completely suppressed after 260 kL. Two faint features can be identified in the sample B spectrum at 260 kL, which nonetheless can be attributed to two energy loss modes of the underlying Ni occurring around 6 and 3.4 eV \cite{Hagelin}. As a confirmation of that, a comparison of this spectrum with one acquired on an empty Ni TEM grid -- same grid type without graphene -- is also shown in Figure \ref{fig:GrTEMH_EELS} (right).

\begin{figure}[h]
\centering
\includegraphics[width=0.65\textwidth,height=0.6\textheight,keepaspectratio,trim={5.5cm 5.5cm 5.5cm 6cm},clip]{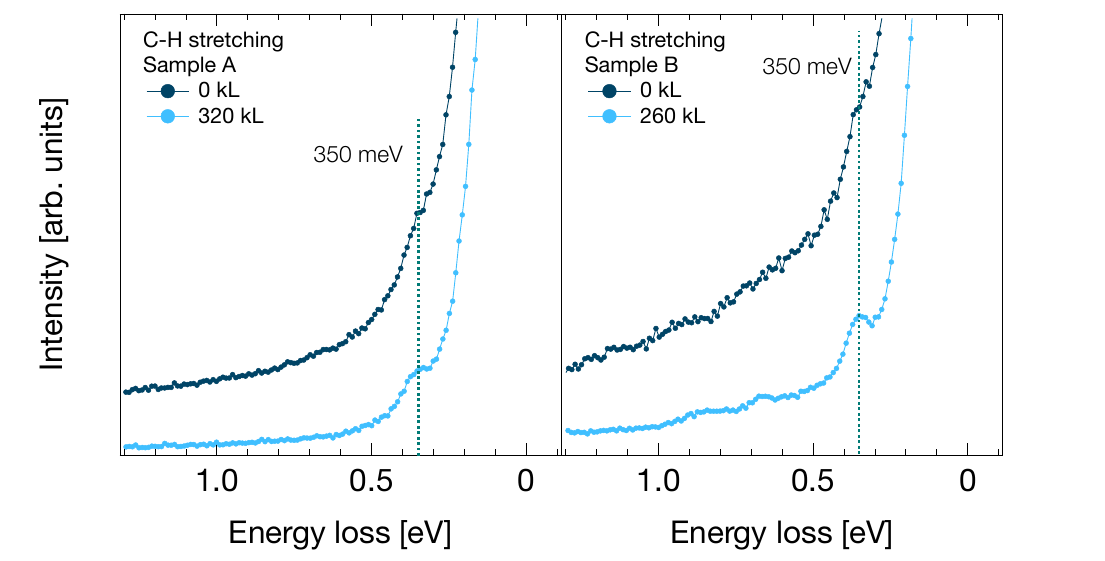} 
\caption{EELS spectra of the vibrational region measured on sample A (left) and B (right) before and after the exposure to atomic hydrogen at saturation (320 kL for sample A and 260 kL for sample B). The spectra before hydrogenation are shown in dark blue and in light blue after exposure. A dashed line at 350 meV is also added as a guide for the eye.}
\label{fig:CH}
\end{figure}

The C-H stretching mode represents a direct evidence of the formation of hydrogen bonding with the C atoms in the graphene lattice. This vibrational mode has an energy of about 350 meV \cite{DiFilippo} and it was observed on both the hydrogenated samples. In Figure \ref{fig:CH}, the EELS spectra focused on the vibrational region close to the elastic peak, before and after hydrogenation saturation, are shown. For both samples, the peak associated to the CH stretching at 350 meV is well visible after hydrogenation indicating the C-H bonds formation after exposure. A feature at the same energy loss can be seen also for the two samples before hydrogenation. The latter probably associated to the $sp^3$ defects of the pristine graphene lattice, which are passivated with hydrogen terminations. Moreover, the spectrum of sample B after hydrogenation shows two additional features at about 660 and 860 meV that can be due to electronic transitions of adsorbed molecules of $\mathrm{C_6H_6}$  and $\mathrm{CH_4}$ respectively.

As already discussed, the distortion of the graphene lattice toward an $sp^3$-like configuration leads to a band gap opening. Being the investigated samples single graphene layers supported by a metallic grid, the measurement of the band gap with optical techniques is non-trivial and it would be dominated by the substrate signal. In addition, they are generally performed in air and the breaking of the ultra-high vacuum condition can lead to the degradation of the sample \cite{stability}. Conversely, EELS is a surface sensitive technique that gives access to the electronic transitions and it is carried out in the same UHV chamber. Thus, by measuring the transition onset in the spectrum, the optical band gap can be obtained. In general, for an insulator or a semiconductor, the elastic peak in the EELS spectrum is followed, after the vibrational range ($\sim$ 40 to 400 meV), by a flat region clear of losses up to the onset of the inelastic electronic excitations. The shape of the onset should be $\propto (E-E_G)^{1/2}$ for direct gap semiconductors and $\propto (E-E_G)^{3/2}$ for indirect gap semiconductors \cite{Rafferty}. 

\begin{figure}[h]
\centering
\includegraphics[width=0.65\textwidth,height=0.6\textheight,keepaspectratio,trim={3cm 4.5cm 3cm 4cm},clip]{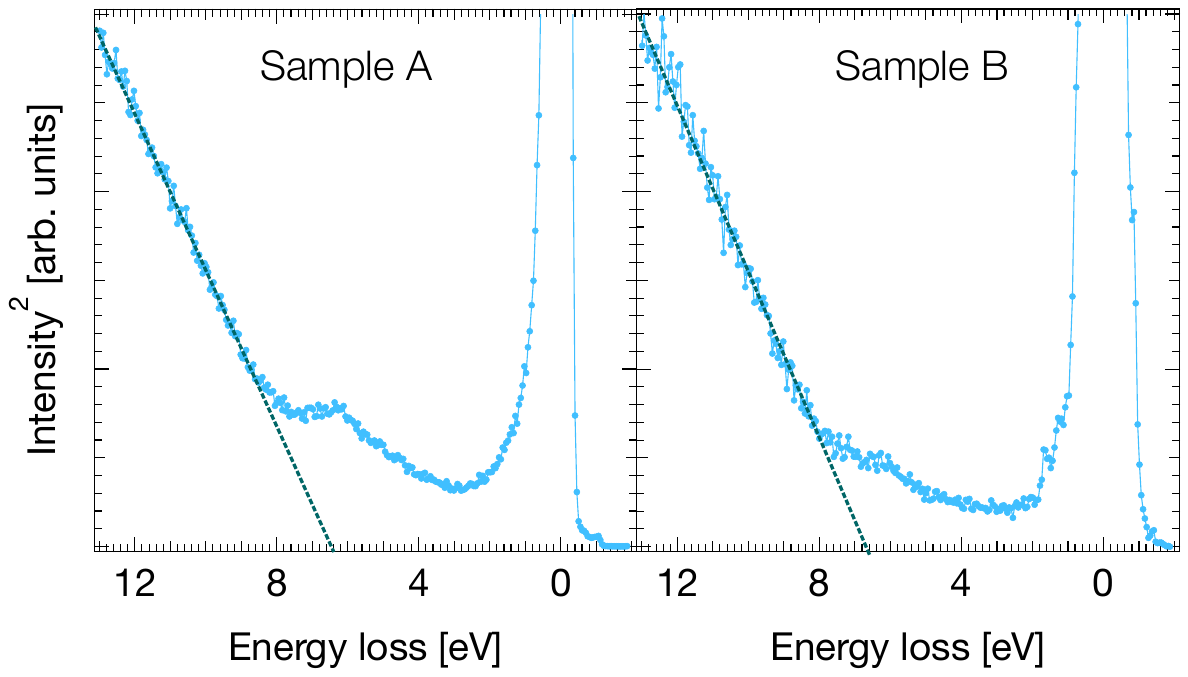} 
\caption{Square of the EELS spectra measured for sample A at 240 kL (left) and sample B at 260 kL (right). In both the plots, the dashed line represents the straight line drawn in the region of the transition onset.}
\label{fig:BandGap}
\end{figure}

The square of the intensity of the EELS spectra of samples A and B after hydrogen exposure is well represented by a straight line in the region of the transition onset, as shown in Figure \ref{fig:BandGap}. This suggests a shape compatible with $(E-E_G)^{1/2}$. Although the onset is hidden underneath the features of Ni and non-hydrogenated graphene, for samples B and A respectively, the zero-intensity baseline can be seen on the right side of the peak. Thus, the interception of the onset straight line with the zero baseline represents the optical gap energy: $E_G$ = 6.2 eV for sample A and $E_G$ = 6.3 eV for sample B. Since no background has been considered underneath the transition onset, the two values of optical band gap obtained with this analysis should be read as lower bounds. Furthermore, the contribution due to excitons would shift the measured gap to lower values as well. 

It is worth noting that the EELS measurements were performed in reflection geometry with an electron beam size of approximately 500 $\mu$m, which is much larger than the grid-hole size ($\sim$6.5 $\mu$m). As a consequence, the measured signal inevitably contains contributions not only from hydrogenated graphene, but also from the underlying nickel grid and, in the case of sample A, from non-hydrogenated graphene regions. In this context, the residual signal in the gap region is expected and the spectral evolution in the 8-12 eV range consistently marks the onset of interband transitions, therefore providing a measurement of the optical band gap.

For a sample of deuterated nanoporous graphene, the optical band gap measured with high resolution EELS (HREELS) can be found in \cite{Betti23}, where the first transition onset is observed at 3.25 eV. The contrast between the results can be ascribed to the different graphene-systems taken into account. Indeed, it has been previously discussed that according to several theoretical predictions the band gap strongly depends on the particular graphene structure and hydrogenation morphology, spanning a wide range of energies (4.7 to 6.1 eV \cite{NPG22, Gao, Lebegue, Cudazzo, Betti23}). However, the opening of a wide gap for both the samples - 6.2 and 6.3 eV - gives a further indication of a two-sided hydrogenation of graphene.

\subsection*{UV photoemission spectroscopy}
A further experimental test to understand the morphology of the C-H bonding is the measurement of the valence band, after the graphene exposure to hydrogen. From the C 1s spectrum, investigated with XPS, it is possible to study whether if the $sp^2$ lattice has been distorted toward an $sp^3$-like configuration of C atoms but it can not be understood how and where the H atoms are bonded to carbon. On the other hand, a good experimental marker may be the valence band measurement via UPS.

\begin{figure}[H]
\centering
\includegraphics[width=0.6\textwidth,height=0.6\textheight,keepaspectratio,trim={4cm 4cm 4cm 4cm},clip]{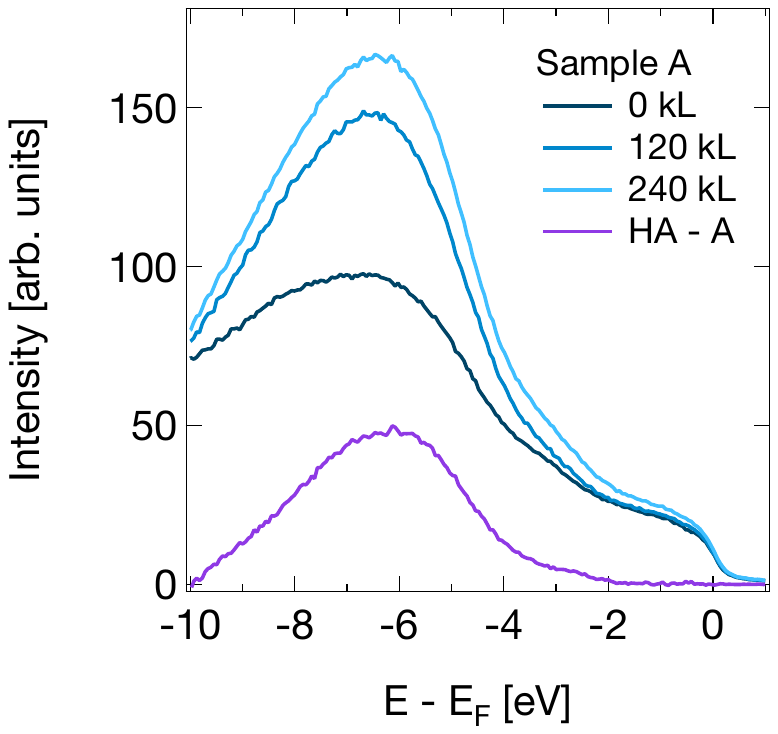} 
\caption{Valence band measured on sample A for increasing H dose (dark to light blue). The violet curve represents the subtraction of the 0 kL spectrum to the 240 kL one, the latter weighted by a factor 0.87.}
\label{fig:VBvsHdose}
\end{figure}

UPS measurements has been carried out on sample A only and in normal emission geometry (\emph{i.e.} in the $\Gamma$ point). In Figure \ref{fig:VBvsHdose}, the valence band evolution of sample A, as a function of the H dose, is shown. In all the spectra the Fermi edge is clearly visible and it is due to the valence band signal of the Ni grid on which the graphene is deposited. Furthermore, since sample A is not completely hydrogenated ($sp^3$ saturation of 59$\%$), the valence band signal after 240 kL will be due to hydrogenated graphene, non-hydrogenated graphene and Ni signal. In order to isolate the hydrogenated contribution, the clean sample curve has been subtracted to the 240 kL one by weighting the latter with a factor 0.87. The weight was chosen by reducing the residual Fermi edge to zero. The subtraction curve is shown in Figure \ref{fig:VBvsHdose} (violet). By taking into account the density of states calculation reported in \cite{NPG22}, the wide peak at around -6 eV and the feature at about -2.8 eV can be ascribed to the coexistence of one-sided and two-sided hydrogenation of the graphene lattice.

\section{Conclusions}
The experiment reported in this work consisted in the thorough characterisation of hydrogenated monolayer graphene on TEM grids with three spectroscopic techniques: XPS, EELS, UPS. The hydrogenation was carried out \emph{in-situ} on two samples with a different starting status, as resulted from XPS. Indeed, the C 1s core-level spectrum revealed a 20$\%$ $sp^3$ contribution in one case and a significantly higher 46$\%$ in the other. After exposure to atomic hydrogen, the latter reached a 100$\%$ $sp^3$-like C 1s profile, while for the first the saturation level was the 62$\%$ even with higher dose. This result is in good agreement with what reported in \cite{Ruffieux}, where the hydrogen adsorption is shown to be favoured by a $sp^3$-like character, resulting in a lower energy barrier for a C atom to be pulled out above the plane defined by its nearest neighbours. Moreover, the binding energy shift of the $sp^3$ component is compatible with a two-sided hydrogenation, but its width does not allow to exclude also a contribution of one-sided regions. Nevertheless, it is worth to underline that the 100$\%$ $sp^3$-saturation of sample B is the highest ever achieved for the hydrogenation of graphene structures.

The EELS measurements confirmed the reduction of the $sp^2$-coordinated carbon atom regions. The $\pi$-plasmon excitation was indeed completely quenched for the 100$\%$ saturated sample and significantly reduced for the other. Furthermore, in both the EELS spectra, the shape of the electronic transition onset was highly compatible with $(E-E_G)^{1/2}$, thus indicating the opening of a wide band gap. A lower bound to the optical band gap of 6.2 eV and 6.3 eV was found for the 62$\%$ and 100$\%$ saturated samples respectively. 

Finally, the valence band spectra measured with UPS on sample A at increasing higher dose, indicate the coexistence of two different morphologies for the C-H bonding in the graphene lattice, if compared with density of states calculations reported in \cite{NPG22}. The spectrum after hydrogenation is indeed compatible with both one-sided and two-sided hydrogenation of monolayer graphene.

To conclude, the three spectroscopic characterisations point in the direction of a strong contribution due to the two-sided hydrogen bonding to graphene -- $sp^3$ shift, wide band gap and valence band. Weaker indications of one-sided hydrogenation were found, suggesting the possible coexistence of the two morphologies.

\section*{Acknowledgements}
We gratefully acknowledge the technical support of Gianfranco Paruzza of the INFN Roma Tre Mechanical Workshop. We are grateful for the financial support from MUR PRIN 2020 'ANDROMeDa' project (Contract No. PRIN 2020Y2JMP5). This project has received funding from the European Union's Horizon 2020 research and innovation programme Graphene Flagship under grant agreement No. 881603. We acknowledge financial support under the National Recovery and Resilience Plan (NRRP), Mission 4, Component 2, Investment 1.1, Call for tender No. 104 published on 2.2.2022 by the Italian Ministry of University and Research (MUR), funded by the European Union -- NextGenerationEU -- Project Title PACE (20227F53E4) -- CUP F53C24000790006 -- Grant Assignment Decree No. 20429 adopted on 6/11/2024 by the Italian Ministry of Ministry of University and Research (MUR). This work was partially supported by The Italian Ministry for Universities and Research (MUR) under the Grant of Excellence Departments, (ARTICOLO 1, COMMI 314--337 LEGGE 232/2016) to Department of Science,
Roma Tre University. We are also grateful for the financial support received by INFN CSN5.

\bibliography{bibliographyGr_Hydrogenation}

\end{document}